\documentclass[12pt,twoside]{article}
\newcommand{\AmS}{{\protect\the\textfont2
  A\kern-.1667em\lower.5ex\hbox{M}\kern-.125emS}}

\title{Einstein And  The Evolving Universe}

\author{A.N.Mitra  \\ 
244 Tagore Park,  Delhi 110 009 India\thanks{e.mail: ganmitra@nde.vsnl.net.in}
 }

\begin{document}

% typeset front matter
\maketitle

\begin{abstract}
A panoramic view,  preceded by a short  background of Newtonian mechanics 
and Maxwellian electrodynamics, 
is offered on the extent of how Einstein's space-time geometry,   
believed to be central to an understanding of  the structure of the universe,  
is overshadowed   by  several  hitherto unheard of  features  like dark matter  
and dark energy,  that  seem  to  be  necessary, but by no means sufficient, 
for a  more   complete  picture. 
\end{abstract}

\section{ Newtonian Mechanics }

Once upon a time there was only Newton with his 3 Laws of Motion. Space 
and time were two distinct and independent entities,  each absolute in its 
own right,   which provided  a joint  playfield for  the activities of Matter,    
(yet another distinct entity)  in accordance with  his 3 Laws of Motion.  
Gravitation was a universal force, again governed by  Newton's diktat,  
which pulled everything far and near,  according to the inverse square law. 
To manage  this huge investment,  Newton  had to take recourse to the tools 
of  Mathematics  for which his own resources proved inadequate however. 
The most important tool in this regard turned out to be Differential Calculus 
which he promptly borrowed from a fellow mathematician Leibnitz. The  
resulting structure  was a beautiful piece of physics clothed in elegant  
mathematics which was Newton's legacy to the world  under the name of 
Classical Mechanics.  It was a most  formidable instrument, capable of   
predicting  the outcome of every type of motion  under the Sun in a fully  
deterministic  manner,  provided only the $initial$ condition was known !  
Relativity as  known today was a far cry at that time, yet the equations of motion  
incorporated Galilean invariance  (which stemmed from Newton's first law).  
\par
This powerful machinery was to rule the world, from terrestrial to the celestial,  
for the next 150 years.  It  proved  so self-sufficient  that  God  had apparently decided
not to have an  explicit role in driving it,  except perhaps watch it  from a distance,  
as a detached  observer !  Indeed this  deterministic  scenario for purely physical  
systems led  DeCartes to enunciate  his celebrated law of Cartesian Partition  
according to which  all $physical$  phenomena were to be totally separated  from  
anything which had to do with  the $psychic$, or  the $mystical$.  God  was however  
not  totally banished from this scenario,  for Newton  had thoughtfully provided for an  
$implicate$ $order$  for the universe as a whole,  whose logic  was best left  "unanswered" .   
During this period,  mathematical thinkers,  and there was a whole galaxy of them \\
$$ [Laplace,  Lagrange, Hamilton ,  Poisson, Fourier ;  Gauss,  Euler, Riemann]  $$  \\
had a field day in shaping and re-shaping  this wonderful machinery to their taste,  
and in the process,  giving  newer and newer  meanings to its physical content.  
In particular,  the "canonical"  Hamiltonian equations of motion  for the time evolution 
of dynamical entities in terms of Poisson brackets,  was a most profound structure  
which (though identical in physical content  to the original form of Newton's laws )  
was later to  prove the "golden road to quantization" at the hands of Dirac. 

\section{Maxwell, Lorentz \& Einstein's Relativity}

In  a totally  different sector of physics,  the  piecemeal  laws of electricity and magnetism  
which had   been building up under different heads (Gauss, Faraday, Biot-Savert) were brought 
together by James Clerk Maxwell  under a single umbrella  of four interlinked differential 
equations  in which his own contribution of Displacement Current proved seminal  for  a 
profound  unification process   giving rise to a consistent  wave theory  wherein   the  
wave velocity turned out to be  precisely  the  velocity of light !  This was another  
masterpiece of   effort to  demonstrate  the  underlying  unity  of the basic laws  of  physics  
despite their outward appearance of  disjointed entities.  As if to drive home  the true 
significance  of this  great result,   H.A. Lorentz  showed that  the Maxwell Equations  were 
$not$ invariant  under the simple  Galilean transformations (the hallmark of the limited relativity 
principle for  Newtonian mechanics),  but rather under a $new$ set of linear transformations 
in which time  and space appeared more symmetrically connected than seen  from  the 
equations of Newtonian  Mechanics. Thus  was born the precursor  of  the  special 
theory of relativity  several decades ahead of its formal inauguration by Einstein. 

\subsection{Special  Theory Of Relativity}

Einstein thus had a two-fold legacy to build on , viz.,  i) Newtonian mechanics and ii) Maxwell's 
electrodynamics,  flanked by two crucial "data", one on the structure of space-time, and  
the other on the possibility of a discrete (quantum) structure of matter. The first emanated 
from the Michelson-Morley experiment pointing to the absence of any ether-like substance 
constituting the vacuum,  while the other stemmed from Max Planck's revolutionary 
explanation of the black-body spectrum in terms of a hitherto unknown constant $h$.   
He took up both challenges in two outstanding papers in a single year--1905--, and 
confirmed both: A) a unified structure of space-time (hitherto thought as two disjointed 
entities); B) corpuscular nature of light (hitherto thought of only as wave). His Special 
Theory of Relativity gave a new meaning to the Lorentz transformations not only through 
the kinematical invariance of a $flat$  space-time entity, but also a more profound result  
at the $dynamical$ level, viz., the formal equivalence  of mass and energy  ($E=m c^2$)
which was to find dramatic manifestations throughout the Twentieth Century in more ways 
than one.  Einstein's active love affairs with Quantum Theory  however ended with his single,  
but  seminal,  paper on the photo-electric effect (which fetched him the Nobel Prize),  while  
the  quantum banner was left to be taken up by other stalwarts (de Broglie, Bose,  Heisenberg, 
Schroedinger,  Dirac, Pauli, Wigner).  For  Einstein, his success with special relativity was 
only a beginning---a sort of appetizer  for  more exciting things in relativity. It was another matter 
that despite being the progenitor of quantum theory, Einstein had profound reservations on  its 
completeness, as evidenced by his dispute with Niels Bohr on the subject.  But except for 
the Einstein-Fock-Podolsky paper, he had little else to offer on  the completeness issue.

\section{General Relativity :  Equivalence Principle }

The year  1905 was truly a landmark year which saw the unification of space-time into a 
single entity at the kinematical level, and a corresponding unification of mass and energy 
at the dynamical level of matter,   both  within the framework of  Special Relativity.  Not  content with  
this big achievement,  Einstein  embarked on the next stage of unification 
(this time of matter with space-time)  by appealing to the $universality$ of gravitation. This 
took him a full decade of mathematical gymnastics, at  the end of which he came up with a 
generalized version of relativity,  one in which space-time is no longer flat but gets curved  
whenever it encounters  the  gravitational attraction of  a lump of matter, the bigger the 
lump the greater the curvature !  This had some  remarkable  logical consequences.  
First, the special status  of `inertial'  frames, viz., ones that move with uniform velocities 
wrt one another  in flat space-time ( characteristics of  special relativity),   gave way 
to a more generalized Equivalence Principle  wherein all frames, including acclerated ones,  
are deemed equivalent to all others. To see  the physical significance of this apparently 
innocuous statement,  consider the famous 
example of  a lift undergoing downward acceleration equal to that of gravity. A man sitting  
in such a  lift will not feel the effect of gravity at all !  Another way to express this result  
is to assert that the gravitational and inertial masses are identical.  This  imbedding of 
gravitation into a curved space-time geometry  has both conceptual and observable  
ramifications.  Conceptual because  in a curved space-time, the line of shortest distance 
(a geodesic) is no longer straight, but $curved$,  the curvature  being  the greater the  
bigger the local mass that  causes  the bending.  Observable because of the possibility  
of   $bending$  of light   in the vicinity of a large mass (such as the Sun),  which was 
dramatically confirmed  during the solar eclipse of 1918 by an expedition led by Arthur Eddington. 
[It was another matter that Eddington had apparently `doctored' some data to suit the theory, 
as revealed  in a  recent book by J. Waller( $Fabulous$ $Science$, Oxford, 2002),  yet it 
was perhaps an irony of fate  that  such  `doctoring' gave a much needed  boost  to the 
Theory (GTR) at its nascent stage of evolution, since  similar subsequent obsevations amply 
bore out  its basic  strength].  Two other observable  consequences that have become 
text-book material are i) the advance of the perihelion of mercury,  and ii) the gravitational  
red-shift of light when it is emitted from a massive body like the Sun. [None 
of these phenomena could however be explained by Newtonian mechanics].  An important  
lesson from these early studies was that GTR once for all transformed space-time from 
its backdrop status in Newtonian mechanics, to  the centre-stage as an active dynamical 
entity  on par with other material objects.  

\section{ GTR And Cosmology} 
 
A dynamical  status of space-time conferred by the new geometry  proved the right  incentive   
for addressing the most important question  concerning  Cosmology  itself, viz., its  
connection with  the structure of the Universe.   To unravel this mystery  needed a continuous 
feedback  between  theory  and observation, of which only one component (theory)  was 
forthcoming  in abundance, while the other ( observation) was to wait for several decades before 
materializing.  In the theory  sector,  the rich structure of the new geometry, with its Riemannian  
metric,  gave rise to  a set of  tensor equations which characterized Einstein's equations, and
proved a field day for mathematicians all over the world -- Friedman, Schwarthschild,  de Sitter
Robertson-Walker,  and later Ray Chaudhury and Vaidya-- to discover newer and newer facets of   
these  tensor equations  emanating from  diverse  types of metrics, employing the most intricate  
techniques of  differential  geometry.  
\par
Among the various solutions, a scenario of great historical interest, and one which has come 
into prominence in the modern era,  concerns the role of the Cosmological term $\Lambda$ 
which Einstein had introduced by hand in his equations for the sake of mathematical consistency.   
For,  in  his attempts to solve these equations for  an idealized $static$ 3-sphere universe filled with 
matter at uniform density,  he realized that the  radius of such  
a universe could not be viewed  as "static" (independent of time) unless there was a counter-term 
to balance the effect of time evolution. But he later abandoned this term as the "biggest blunder 
in his life".  de Sitter (1917)  on the other hand,  picked up this item  where Einstein had left it,  
by  recognizing, a la Einstein,  its need  for balancing the gravitational attraction of matter,  so  
as to  produce  a  static universe  where the mean density of matter, and the mean curvature of 
space  would stay  constant.  de Sitter then observed that he could obtain another static model  
by removing all the matter from the original  Einstein model, but now the (repulsive) $\Lambda$-term  
would cause test  particles to accelerate away from each other.  The rate of  this separation  was  
predicted  by H. Weyl (1923) to follow the simple law $ v = H. (distance)$.  Similar derivations  of 
an evolving universe with the same law of separation were also given by A. Friedmann (1922)  
and   G. Lemaitre (1927), which was experimentally confirmed by  Hubble (1929).  
\par    
In a landmark theoretical development, George Gamow (1946) proposed that matter in the  
early universe was dense enough to undergo rapid thermonuclear reaction, and that energy  
densities were radiation-dominated. Soon afterwards,  R. Alpher, H.Bethe and G. Gamow (1948)
predicted that the black-body radiation that originally filled the universe, should have a Planck 
spectrum corresponding to a temperature of about $25^0 K$. [This was the famous " $\alpha\beta\gamma$ 
paper"   put in without Bethe's formal consent (!); so when the theory was later in  (temporary) 
trouble, Bethe had alledgedly  wished   his  name were Zacharias !] .  The eventual observation 
by A. Penzias and R. Wilson (1964) of an unexpected  background radiation of 7 cm,  with a  
temperature of about $3.5^0 K$,  and  its immediate identification by  Dicke-Peebles- 
Roll-Wilkinson as the expected relic radiation (a la the $\alpha\beta\gamma$ 
  paper), was the first 
major experimental confirmation of the "Big-Bang" scenario.  [A parallel proposal by 
Bondi-Gold-Hoyle (1948),  later to be known as the Hoyle-Narlikar Steady State Theory, 
had to be abandoned in response to the Penzias-Wilson discovery].   
\par
An important  prediction by the Indian astrophysicist Subramaniam Chandrasekhar,    
of the  existence of a   critical mass -- the Chandrasekhar limit -- beyond which the star collapses 
under its weight ,   met with  stiff resistance from Eddington,  but the profound nature of the 
discovery eventually  fetched him   the Nobel Prize.    Other  
outstanding predictions  of these investigations  included  i) gravitational radiation, ii) the expanding  
universe, and iii) black holes as the final  stage of  dense neutron stars.   This is about as far as 
Einstein's GTR machinery could go towards unravelling the mysteries of Cosmology, taking into  
account the severe experimental limitations  of the time,  but more tests were in the offing.  

\subsection{ Experimental Discoveries: Pulsars; GPS }

Towards the end of the last century, great strides in high precision instrumentation, and in the observational  
techniques of astronomy, have led to new precision tests of GTR predictions. Over a thousand neutron  
stars have been found in the form of pulsars (fast rotating stars) whose gravitational fields can be  
adequately described in terms of GTR only.  Another  important discovery was  a 
binary pulsar [R. Hulse and J. Taylor, 1976]  whose orbital motion  measurement with great  accuracy, 
led to a  precision test of GTR.  Still  another important observation  was that as a result 
of the emission  of energy into gravitational radiation, the total energy of the orbital motion 
$decreases$ with time  at a rate predicted by GTR to within a third of a percent.  This has led to a standard GTR 
correction to the flow of time on orbiting satellites, as compared to the corresponding rate on earth,  
as an essential part of the Global Positioning System (GPS), which allows various users (commercial, 
military, etc) to calculate a precise location on the surface of the earth, and to transfer accurate time  
readings using triangulation with satellite signals. 

\subsubsection{Black Holes; Quasers}

In the views of S. Chandrasekhar ( as elaborated by  the  famous GTR specialist Abhay Ashtekar in the 
Indian Acad Sci publication Patrika, March 2005),  " black holes of nature are the most perfect macroscopic 
objects in the universe,  the only elements in their construction being concepts of space and time ".  Black  
holes have also  proved  a gold-mine for generating ideas on fundalmental physics. Indeed their amazing 
variety of properties have intrigued quantum field theorists and relativists alike, and provided insights  
into the inter-connection between general relativity, quantum theory and  statistical physics,  which 
constitute the three pillars of modern physics.   
\par
Many black hole candidates have been identified via antronomical observations. They can be broadly  
classified  under two heads : i) those arising from the collapse of stars, having masses of the order of  
$1-10$ times that of the Sun, and radii of a few kilometers; ii) those found  at  the centres of galaxies, 
having masses of the order of millions to billions of times that of the Sun, and radii comparable to that  
of the solar system. Our own galaxy may well contain such a black hole ! Not only that, the most violently 
energetic  objects in the universe--the $quasers$-- are thought to be powered by accretion of matter onto  
such huge spinning black holes. 

\subsection{ Expanding Universe}

One of the most dramatic predictions of GTR is the theory of the expanding universe which has been 
convincingly  confirmed by observation of the  velocities of distant objects. The gravitational red shift  
of spectral lines, which was initially a most difficult  test of GTR,  has now become a standard tool of 
astronomy.  In the same way, the bending of light by the Sun, is now a routine technique to map dark 
matter using  gravitational lensing.  Indeed the mass of intervening galaxies is often observed to  
distort the light   from more distant sources quite significantly,  resulting in the production of multiple 
images. This provides   a  method  for searching  massive objects that produce no detectable radiation.  
All this is in complete  accord with the predictions of GTR.   

\section{ GTR-QFT Unification Issues}

All this  constitutes very impressive confirmation of the basic tenets of GTR, yet there are compelling  
reasons to  believe that  there are still unknown facets to gravity than  are contained in these results. 
The biggest  goal  is now the need for  a consistent theory that comprises both GTR and quantum 
theory  (QFT).  Now  the degree of unification  achieved within GTR  has   been  outlined in Sects 3-4.   
And in the QFT sector,  the  unification achieved so far is equally impressive. This last   
is symbolized by   Dirac's synthesis of quantum mechanics and  special relativity  which together 
have resulted in the  prediction  of  antimatter.  Indeed  quantum  theory  has  even covered  the problem  
of   interaction of radiation with matter --  QED that is--  by  addressing  the problem  $virtual$  processes ( the 
problem of emission and subsequent absorption of radiation)  which  was  fraught with dangerous infinities 
that  would not make sense for $physical$ processes !  The solution lay in the absorption of  infinities 
through  a redefinition of physical entities like mass and charge (in terms of `bare' charge and mass),  a process  
termed  $Renormalization$, so that physical process could be expressed entirely in terms of the 
`renormalized'  quantities only.  A consistent treatment further required that the operation be independent 
of the inertial frame under consideration. This was eventually achieved by the Covariant  QED Formalism  
of Tomonaga-Schwinger-Feynman-Dyson, a truly  great theory which achieved experimental 
confirmation to within one part in a trillion ! 

\subsection{Hawking Paradox}

A  far bigger challenge  at this stage is  the unification of GTR and quantum theory as  this goal 
is fraught with major  conceptual  problems. The  nature of such conceptual problems is best 
illustrated  by   Stephen Hawking's discovery that when the effects of quantum mechanics  
are included, black holes start emitting radiation,  i.e., they are no longer black ! Of course this 
radiation is too small to be detectable, but it is conceptually very important. In Hawking's approximate 
calculation, however, this radiation is randomly distributed, i.e., in a $thermal$ manner. It is not apriori  
clear whether  this is an exact result or not,  but if so,  then  the  causal connection between the past and  
the future-- an essential characteristic of quantum mechanics--gets lost.  In other words,  a  black hole 
which, in principle, carries quantum information from the past  emanating  from the objects  it has  
swallowed in the past, can evaporate by radiating in a randum  fashion, thus  apparently  violating the 
law of causality.  This is the famous Hawking paradox whose   ultimate resolution may well  be a key to  
the understanding  of  the quantum nature of  space-time. 

\subsection{Initial Conditions, Etc}

Then comes the question of  initial conditions  which are left unanswered in the essential framework that 
GTR provides for the understanding of Big Bang Cosmology. To give a simple analogy, Newton's theory 
describes the motion of planets to be sure, but does $not$ determine the size and shape of the solar system, 
which in turn would have needed the specific details of its history. Again,  other sectors of the universe 
have different features from ours.  Yet the universe as a whole  has some strikingly simple  features  
like \textit{approximate homogeneity} and \textit{spatial flatness}, which a fuller theory is expected to explain. 
Homogeneity means basically  that any large region of the universe of a given age looks much like any  
other region of the same age. Spatial flatness means that $space$ by itself (not space-time !) is $flat$ 
on large scales. Both these properties have been observed and measured with considerable precision, 
through studies of the micro-wave background radiation. Note that neither homogeneity nor spatial  
flatness are required by  by classical GTR  but are  at least allowed by it.    Now questions  like 
"Why is our universe  so homogeneous and flat ? ",   call for some extra ingredient  beyond the 
premises of GTR,  ingredients  which  are no less concerned  with  the ramifications  of   
quantum theory (QFT) down to  the earliest moments of the Big Bang. This in turn would require the 
calculation of the behaviour of quantum gravity at high energies, something which is not known 
at the moment.  Stated differently,  a synthesis  of  GTR (the theory of space-time) with  QFT 
requires  the introduction of ideas which impinge on both sectors in a highly interlinked manner, 
consistently with the new observational features of Cosmology.  

\subsection{ Unification Candidates: Inflation}

One such idea,  which has been highly successful, is $inflation$ first proposed by A. Guth.   He  assumed 
that the universe, early in its history underwent a period of exceptionally rapid expansion. 
Now expansion tends to decrease spatial curvature, just as the blowing up  of a baloon  
makes its  surface appear flatter. The enormous expansion  associated with inflation  means  
that  the universe we see today began from a very tiny region of space that could have been 
smooth before inflation. While inflation cannot fully eliminate the dependence of the state of 
the universe today upon its initial state, it can at least considerably reduce that effect.    
A great advantage of the inflation theory is that it  is rooted in concepts associated with the   
particle  physics scenario, thus fulfilling the condition of synthesis of GTR  with   QFT. 

\subsection{ Condensates In QFT: Dark Matter}

 Now  unified theories   of particle physics, in turn,  require the existence of $condensates$  which  
are   the relics of  "symmetry breaking"  (spontaneous or otherwise) in the theoretical framework.  
Even  without going   into the details of the symmetry-breaking mechanism,  the immediate 
consequence of the  existence of these condensates is that  their very presence 
indicates that   the symmetry  of the fundamental equations  are "broken",  as  demanded   
from  observation.  Now since these  condensates  belong to the "matter part"  of 
the GTR equations, they must be compatible with the observation that visible matter  is  
only about 5 percent of the total amount of matter needed to account for the consistency 
of these equations. Therefore the rest must be invisible or $dark$ matter,  whose  identity   
is  thus one of the key questions of GTR cosmology  today.  While the weakly interacting  
neutrinos  by virtue of their neutral charge,  are an ideal candidate for dark matter,  their 
negligible masses and light-like velocities  are  impedimemts  in the way  of  accounting  
for the (95 percent)   dark matter, as is is hard to see how  they would be  gravitationally 
trapped  in  density fluctuations in the early universe.  A more promising  candidate  which  
is  compatible with this cosmological requirement,  is   the  (heavy)  $neutralino$  which 
is a  new electrically neutral stable particle arising  from the breaking of  supersymmetry,  
but which interacts very  weakly with matter.  [It is yet to be experimentally identified,  
despite several ideas for its detection in high energy accelerator based experiments].  
Another hypothetical candidate -- the $axion$ -- was introduced by Weinberg in the 
strong interaction (QCD) sector of the Standard Model to compensate  for the observed lack 
of  CP (charge-parity  or matter-antimatter symmetry ) violation, since unfortunately QCD  
would otherwise predict  a small CP - violating phase ($\theta$)  in its structure.  The $axion$ 
is  thus predicted as an  additional  low mass weakly interacting  particle which would have  
been produced abundantly during the Big Bang and thus could easily account for the 
needed amount of dark matter.  Other dark matter candidates have also been suggested,  
but no final solution has yet been found.  Understanding the nature of dark matter  is today
one of the  most challenging problems  for  the  unification of matter with space-time.  

\subsection{ Temperature Dependence of Condensates}

Now  to see how the behaviour of condensates  with increase in temperature holds the 
key to  an understanding of  $inflation$, we need to go through a twin logic :   i) standard phase  
transition associated with a condensate when a `broken symmetry'  is restored ; ii) the 
gravitational behaviour of the energy  when  it is trapped in the  condensate, versus when 
it was `free'  at the higher temperature. (i)  In a standard phase transition, when the 
temperature is raised sufficiently, the  condensate just evaporates away , (like the melting  
of ice   into water).  Stated differently,   the broken $symmetries$  associated  with  the 
condensates  are sort of "restored".  This   may be seen  by  analogy with the behaviour 
of  an  ordinary magnet where, at low  temperatures, the spins are aligned  in  some  
preferred  direction,  since such a  configuration is energetically favourable. However,  
as the  temperature  is increased,  such an alignment is no longer energetically favourable,  
and the configuration tends  to be   `isotropic'  (more symmetrical), with the spin 
directions getting more randomly oriented.  (ii) The second part of the logic  concerns the 
rather  different behaviour of the \textit{vacuum--energy}   as it undergoes a  transition  from the 
higher (no condensate)  to the lower (condensate) temperature state.  Now the lower  
temperature corresponds to the situation  where  the condensate  contains an enormous 
amount of \textit{vacuum--energy} in a "trapped "  form , obeying the `normal' laws of  
gravitation. On the other hand, at the higher temperature, the same \textit{vacuum--energy} 
has entirely different gravitational properties from its  (more conventional)  
"trapped form" .  Namely, if the vacuum energy is dissipated "slowly" , it causes an 
exponentially rapid $expansion$ of the universe,  giving  rise to  a period of $inflation$.    
            
\subsection{ Baryon Asymmetry: CP Violation}

Another  observational aspect of the  universe is the preponderance of matter over antimatter.  
Now in the  Standard Model, this preponderance is totally absent, i.e.,  the number of baryons 
minus the number of anti-baryons is strictly conserved. Now if this principle is accepted 
for all time, then the observed  baryon asymmetry would be merely an initial condition, 
a legacy of the original big bang.  On the other hand, in $unified$ theories,  the baryon  
number evolves with time (since quarks can change to anti quarks and / or other 
particles), leading to exotic phenomena like proton decay ! If such processes do indeed  
occur, then (as explained in the preceding paragraph), the symmetry  would be  valid  at 
sufficiently high  temparatures,  down to the moment of the big bang.  In this alternative 
scenario, the present preponderance of matter over anti-matter must be regarded as 
the result of  cosmological evolution of the equations of motion.  
 
\subsection{Accelerated Expansion: Dark Energy} 

Finally, any theory of space-time-matter must address the most mysterious  
question in physical science: the nature of the $vacuum$  which is believed to  
be populated with the virtual particles on the one hand, and the symmetry-breaking  
condensates on the other. The definition of zero energy can be arbitrarily adjusted  
in many theories, but once adjusted in one epoch, it cannot be altered, and the  
effect of quantum corrections must then give the vacuum energy in $all$ $epochs$.  
To set   the question of   vacuum energy  in the GTR language of Sect.4,  it is  
tempting to identify  this quantity  with  the energy  of  a de Sitter universe  with  
positive cosmological  constant $\Lambda$.  [ A negative $\Lambda$ would  
correspond to  an anti-de Sitter or "AdS"  vacuum].  A quantitative connection  is  
however  fraught with  danger, since the quantum corrections give values of the 
expected scale of energies far in excess of what is allowed  experimentally !  
Indeed,   the discrepancy is so large ($\sim 10^{55}$) that  it indicates a big gap in 
the understanding of vacuum  in terms of  gravity !    
\par
Now  the most intriguing hint of a new physics from cosmology  is the $observation$ that  
the expansion of the universe is {\it speeding up}, rather than slowing down, thus implying   
the presence of a mysterious form of energy,   termed  {\it dark--energy},  that pervades the 
universe with a gravitational effect  that is $repulsive$  rather than attractive.  Indeed the 
latest observation suggests that this energy which corresponds to $negative$  pressure,  
constitutes about  70  percent of the energy density of the universe, the other 30 percent 
being shared between visible matter ($5$) and dark matter ($25$).   
The observational basis  for such a  composition of the universe  
comes from the  study  of temperature anisotropies in the cosmic background temperature  
radiation (CMBR),  which (as noted in Sect. 4),  is a relic  from the hotter  phase 
of the universe when  it was about a thousand times smaller in size !         
\par
We have by  now encountered three kinds of energy: i)  the (observed) dark energy; ii)  the 
(quantum) energy  of the vacuum;   and iii) the (GTR) motivated cosmological constant  
of the de Sitter universe.  They are presumably interconnected, but so far there is 
no deeper theory  for a natural understanding of these apparently disparate items, 
taking account of the ubiquitous role  of gravitation.  

\section{ Strings To The Rescue ?}

In the quest for a unified  quantum theory  which includes gravitation,  a natural direction  
to look for  is the only available candidate on the horizon, viz.,  the  3 decade old  String Theory 
which was motivated by precisely  such a need,  and has undergone  several stages  
of sophistication ,  from  string theory  to  superstring theory,   and now  to  $M$-theory. 
Unfortunately  no specifically testable prediction of this grand theory has emerged 
so far,  yet many believe that the theory is probably on the right track.  As to  its   
main  features,  string theory  is concerned with the problem of constructing a  
consistent quantum theory  of "extended particles" or strings, and $not$ point 
particles to start with.  Its most startling consequence is the prediction of  the  
existence of gravity within its basic framework, with the added advantage that  
(unlike conventional GTR), it does not suffer from the problem of infinite quantum  
corrections. [This is  not  as  mysterious  as might  appear on  first  sight, since 
a  certain $length$ dimension associated with this extended object  is available for  
"toning" down the effect of an otherwise naked infinity associated with $piont$  
structures and their interactions].  Unfortunately  a consistent string theory  does  
not work in 4 dimensions, the minimum number of dimensions needed for a  
consistent description being $eleven$.  Therefore it is necessary to assume that  
these additional (7) dimensions get curled up, leaving the usual 4 (extended)  
dimensions to play with.  The new theory yields many solutions,  some with   
the known features of the existing theories ( unification of couplings, supersymmetry, 
and axions), but there are other equally consistent solutions that lack  these  
features. So far there is no reason to prefer one solution to another, nor has a  
unique solution emerged which accounts for all observations.  Supersymmetry  
which is a key feature of String Theory,   predicts a $doubling$ of the number of  
known particles, and although this prediction has not yet  been fulfilled, the  
predicted particles  are believed to be well within the range of the  next  
generation of  particle accelerators.  
\par
A key question now concerns the possibility to understand the twin features of 
inflation and an accelerating universe ( de Sitter universe with positive $\Lambda$) 
within the ambit of string theory. Recent progress in this direction [ see  for a  review : 
S.P. Trivedi, Curr. Sci. vol 88,  p 1125 (2005) ] seems to suggest  that,  
despite the existence of  various no-go theorems in the way of such pursuits, 
alternative scenarios are available  within the broad framework of  string theory  
for the construction of  de Sitter universes with the  desired properties which  
may  be able to circumvent the no-go theorems.  Similar possibilities also   seem  to  
exist   for the understanding of inflation within the same broad framework.  The details  
are however too technical to warrant elaboration. 
\par
Although no explicit references are given, I wish to acknowledge that I have greatly 
benefitted from two key references: i) Gravitation  by Meissner et al, considered as the 
" Bible " ; ii) Connecting Quarks with the Cosmos, National Acadeies Press,  
Washington, D.C., 2003.  From both these classics, I have frequently drawn ideas 
in sequence (albeit in my own language),  but   without explicit reference in context.

\end{document}